\documentclass[aps,prl,english,showpacs,twocolumn]{revtex4}
\usepackage{latexsym}
\usepackage{bm}
\usepackage{babel}
\usepackage[T1]{fontenc}
\usepackage[latin1]{inputenc}
\usepackage{mathrsfs}
\usepackage{amsmath, amssymb}
\usepackage{graphicx}
\usepackage{subfigure}

\begin{document}
\title{Pyramidal Atoms: Berylliumlike Hollow States. }
\author{Marianne Dahlerup Poulsen}
\author{Lars Bojer Madsen}
\affiliation{Department of Physics and Astronomy,
  University of Aarhus, 8000 {\AA}rhus C, Denmark}

\begin{abstract}
Based on the idea that four excited electrons arrange themselves
around the nucleus in the corners of a pyramid in order to
minimize their mutual repulsion, we present an analytical model of
quadruply excited states. The model shows excellent comparison
with {\it ab initio} results and provides a clear physical picture
of the intrinsic motion of the four electrons. The model is used
to predict configuration-mixing fractions and spectra of these
highly correlated states.
\end{abstract}

\maketitle Describing electron correlations in terms of simple
physical pictures requires an understanding of the
inter-electronic interaction which goes beyond numerical
diagonalization of the Hamiltonian of the system. The study of
electron-electron correlation in multiply excited states has
contributed a lot to this understanding and in the eighties group
theoretical~\cite{Herrick83,Rau90}, molecular
adiabatic~\cite{Feagin86}, and hyperspherical~\cite{Lin84}
approaches led to an understanding of doubly excited states in
terms of approximate quantum numbers associated with angular and
radial correlations. In the nineties advances in light sources
stimulated research on triply excited hollow states in Li (see
Ref.~\cite{Madsen03} for a recent review). In particular,
theoretical models referred to as the three-electron ionization
ladder~\cite{Komninos88,Nicolaides90}, the normal mode
analysis~\cite{Watanabe87,Bao97,Morishita01b}, and the symmetric
rotor~\cite{MMPRL,MMPRA,MMJPB} confirmed the appealing physical
picture for the lowest-lying triply excited states that the three
electrons be arranged in an equilateral triangle coplanar with the
nucleus situated in the center, and identified the magnetic
quantum number with respect to the body-fixed $z$ axis as
approximately conserved.

Current interest in multiply excited states is strongly stimulated
by further development of synchrotron radiation sources and
free-electron lasers providing unprecedented brightness in the uv
and xuv and therefore readily access to the spectral region of
interest. Successful access to hollow lithium
states~\cite{Madsen03} makes studies of quadruply excited states a
natural next step in this development. Sofar there are no
experiments and only very few theoretical studies on these
systems. One work~\cite{KN} considered the lowest-lying quadruply
excited $^5$S$^\text{o}$ state in Be by multiconfigurational
Hartree-Fock calculations while two other works~\cite{Bao96,Bao98}
presented a classification scheme based on symmetry
considerations. Hence, in view of experimental advances, it seems
timely to aim at a better understanding in the form of physical
images, including geometry, symmetry considerations and electronic
motion for highly correlated four-electron systems. Here we shall
concentrate on intra-shell states with the electrons confined to
the same shell. These states are interesting since the dominant
configurations of the lowest-lying quadruply excited berylliumlike
states are intra-shell states.
\begin{figure}[htb] \centering
\includegraphics[width=0.9\columnwidth]{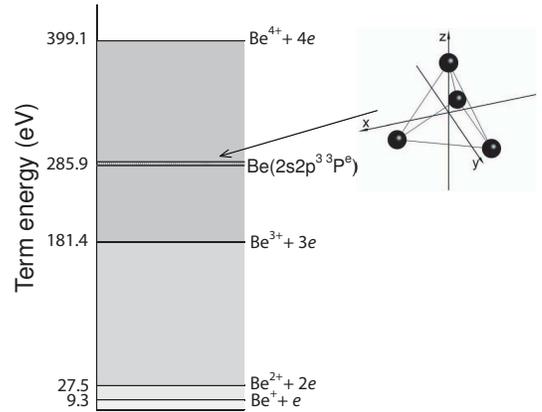}
\caption{Thresholds and the lowest quadruply excited
$^3$P$^\text{e}$ state in Be. The insert to the right shows, in
the body-fixed frame, the physical picture on which our model is
based. The nucleus is located in the origin and the circles denote
the electrons.}\label{illu4}
\end{figure}

Figure \ref{illu4} shows the thresholds in Be, the lowest
$^3$P$^\text{e}$ quadruply excited resonance state and an insert
of the physical picture on which our model is based. We note that
while two and three electrons are confined to a line and a plane,
respectively, the extent of four randomly placed electrons is
three dimensional. Furthermore, the electron-electron correlations
are expected to be stronger when more electrons interact.
Therefore, research on these states opens new possibilities to
test theoretical approximations for correlated electron motion.

We consider the Coulombic system of four electrons moving in the
field of a point-like nucleus of charge $Z$ corresponding to the
nonrelativistic Hamiltonian
\begin{equation}\label{Hamil}
\hat{H}=\sum_{j=1}^4\Big(\frac{-\hbar^2}{2\mu}\nabla_j^2-\frac{Ze^2}{r_j}\Big)+
\sum_{i<j}\frac{e^2}{|\vec{r_i}-\vec{r_j}|}\;,
\end{equation}
where $\mu$ is the reduced electron mass and $e$ is the electron
charge. The electron-electron interaction complicates the problem
since the potential is not diagonal in a basis of antisymmetrized
products of single-electron configurations. Consequently,
single-electron quantum numbers are not conserved and the
eigenfunctions are represented as superpositions of
configurations. Our model accounts for the electron-electron
interaction by assuming the wave function be a product of
carefully chosen one-electron states. Furthermore, the ansatz for
the wave function must be supplemented to take into account basic
quantum symmetries (parity, spin, etc.)

The Coulomb repulsion between the four electrons is minimized when
they are placed in the vertices of a regular tetrahedron (see
Fig.~\ref{illu4}). In this configuration the interaction between
any electron and the remaining three is represented by a repulsive
electric field along the direction, $\xi$, of the electron
considered with respect to the nucleus. Hence, the Hamiltonian in
Eq.~\eqref{Hamil} can be approximated by a sum of hydrogenic
Hamiltonians perturbed by weak external electric fields causing
only intra-shell transitions. The eigenstates of the hydrogenic
Hamiltonian perturbed by a weak external electric field are the
polarized Stark states, $|nkm\rangle$, described by the principal,
parabolic and azimuthal quantum numbers denoted $n, k$ and $m$,
respectively~\cite{Baugh97}. The wave function therefore splits up
into a product of one-electron states which is the simplest
possible way to represent a many-particle wave function. Thus, the
basic idea in the model is that the inter-electronic interaction
tends to stabilize the electrons in individual Stark states (see
Refs.~\cite{Drukarev82,Moelmer88} for two and \cite{MMPRL} for
three electron cases). The parabolic quantum number, $k$, may be
associated with the electric dipole moment of the atom through the
relation $p_z=\frac{3}{2Z}enk$~\cite{Baugh97}. Hence, the largest
energy shift and thus the lowest-lying state for each $n$ is
obtained when the Stark states are maximally polarized along
$\xi_i$, $|n,k=n-1,m=0\rangle_{\xi_i}$. The ansatz for the wave
function is therefore
\begin{equation}\label{ansatz4}
|\Upsilon_R\rangle=|nn-10\rangle_{\xi_1}|nn-10\rangle_{\xi_2}|nn-10\rangle_{\xi_3}|nn-10\rangle_{\xi_4}\;,
\end{equation}
with $\xi_j$ the direction of the $j$th electron with respect to
the nucleus. Quantum numbers in kets without a subscript refer to
quantization along the $z$ axis so the electron on this axis is
represented by the state $|nn-10\rangle$. The other three states,
$|nn-10\rangle_{\xi_q}, \,q=1,2,3$, are related to it via
rotations
$|nn-10\rangle_{\xi_q}=\textbf{R}\big(q\frac{2\pi}{3},\Phi,0\big)|nn-10\rangle
=e^{-iq\frac{2\pi}{3}\hat{L}_z}e^{-i\Phi\hat{L}_y}|nn-10\rangle\;,$
with $\Phi=109.5^{\text{o}}$ the inter-electronic angle as seen
from the nucleus (see Fig.~\ref{illu4}).

The wave function $|\Upsilon_R\rangle$ is anisotropic since the
electrons are oriented. As a consequence of this break of
rotational symmetry, $|\Upsilon_R\rangle$ is not an eigenstate of
\textbf{$\hat{L}^2$} and \textbf{$\hat{L}_z$}. Contrary, the
rotational average over Euler angles $\omega
=(\alpha,\beta,\gamma)$
\begin{equation}\label{rotor1}
|\Upsilon_{R}^{LMM_I \pi}\rangle=
\int_{0}^{2\pi}\!\!\!
d\alpha\int_{0}^{2\pi}\!\!\!d\gamma\int_{0}^{\pi}\!\!\!d\beta\sin\beta
D_{MM_I}^{L}(\omega)^{*}|\Upsilon_R\rangle_\omega\;,
\end{equation}
is so~\cite{MMPRA}. Here $\pi$ denotes the parity and $M_I$, is
the magnetic quantum number with respect to the body-fixed $z$
axis. In molecular theory this is an important quantum number
specifying the rotational state of the molecule. Since the
Hamiltonian is spin-independent the total wave function is simply
a pro\-duct of the spatial function,
$|\Upsilon_{R}^{LMM_I\pi}\rangle$, and a spin function,
$|\chi_{M_s}^{S_{12}S_{34}S}\rangle$. The spin function specifies
the total spin, $S$, of the four electrons and its projection,
$M_s$, which are both conserved quantities. The indices, $S_{12}$
and $S_{34}$, refer to the intermediate spins arising from
coupling the spin of the first and second  and the third and
fourth electron, respectively. The total wave function is obtained
by antisymmetrizing the product state
$|\Upsilon_{R}^{LMM_I\pi}\rangle|\chi_{M_s}^{S_{12}S_{34}S}\rangle$,
\begin{equation}\label{anti4}
|\Psi_{R,S_{12}S_{34}}^{LMM_I\pi,SM_s}\rangle=\textbf{\emph{A}}\big[|\Upsilon_{R}^{LMM_I\pi}\rangle|\chi_{M_s}^{S_{12}S_{34}S}\rangle\big]\;,
\end{equation}
where \textbf{\emph{A}} is the four-particle antisymmetrization
operator. The final expression for the four-electron state is
$|\Psi_{R}^{LMM_I\pi,SM_s}\rangle=\emph{N}\sum_{i=\{S_{12}S_{34}\}}|\Psi_{R,i}^{LMM_I\pi,SM_s}\rangle\;,$
with \emph{N} being a normalization constant. The possible values
of intermediate spins, $S_{12}$ and $S_{34}$, are determined by
$\vec{S}=\vec{S}_{12}+\vec{S}_{34}$ added as angular momentum
operators.
\begin{table*}
  \centering
  \caption{Configuration-mixing fractions in Be for the most important intra-shell
  configuration states of ${}^5\text{S}^{\text{o}}$ symmetry.
  SR denotes predictions within the present symmetric rotor model, and MCHF denotes
 multiconfigurational Hartree-Fock calculations~\cite{KN}.}
 \begin{ruledtabular}
 \begin{tabular}{cllllllll}
   ${}^5\text{S}^{\text{o}}$ & \multicolumn{2}{c}{n=3\phantom{00}} & \multicolumn{2}{c}{n=4\phantom{00}} & \multicolumn{2}{c}{n=5\phantom{00}} & \multicolumn{2}{c}{n=6\phantom{00}}\\
   \hline
   Configuration & \multicolumn{1}{c}{SR} & \multicolumn{1}{c}{MCHF} & \multicolumn{1}{c}{SR\phantom{0}}
   & \multicolumn{1}{c}{MCHF} & \multicolumn{1}{c}{SR\phantom{0}} & \multicolumn{1}{c}{MCHF}
   & \multicolumn{1}{c}{SR\phantom{0}} & \multicolumn{1}{c}{MCHF}\\
   \hline
   sp$^3$ & 0.94 & \phantom{$<$}0.90\phantom{00} & 0.81 & \phantom{$<$}0.76\phantom{00} & 0.67 & \phantom{$<$}0.66\phantom{00} & 0.55 & \phantom{$<$}0.59\phantom{00}\\
   spd$^2$ & 0.063 & \phantom{$<$}0.084 & 0.15 & \phantom{$<$}0.18 & 0.20 & \phantom{$<$}0.21 & 0.23 & \phantom{$<$}0.23\\
   sd$^2$f &  &  & 0.013 & \phantom{$<$}0.0081 & 0.039 & \phantom{$<$}0.040 & 0.068 & \phantom{$<$}0.053\\
   p$^2$df &  &  & 0.022 & \phantom{$<$}0.032 & 0.057 & \phantom{$<$}0.058 & 0.091 & \phantom{$<$}0.078\\
   pdf$^2$ &  &  & 0.0014 & \phantom{$<$}0.0025 & 0.0085 & \phantom{$<$}0.0081 & 0.021 & \phantom{$<$}0.017\\
   d$^3$f &  &  & 0.0022 & \phantom{$<$}0.0025 & 0.0097 & \phantom{$<$}0.0064 & 0.021 & \phantom{$<$}0.012\\
    \end{tabular}
 \end{ruledtabular}
  \label{5S}
\end{table*}

\begin{figure}
    \begin{center}
    \begin{tabular}{l}
    \subfigure[]{\includegraphics[scale=.4]{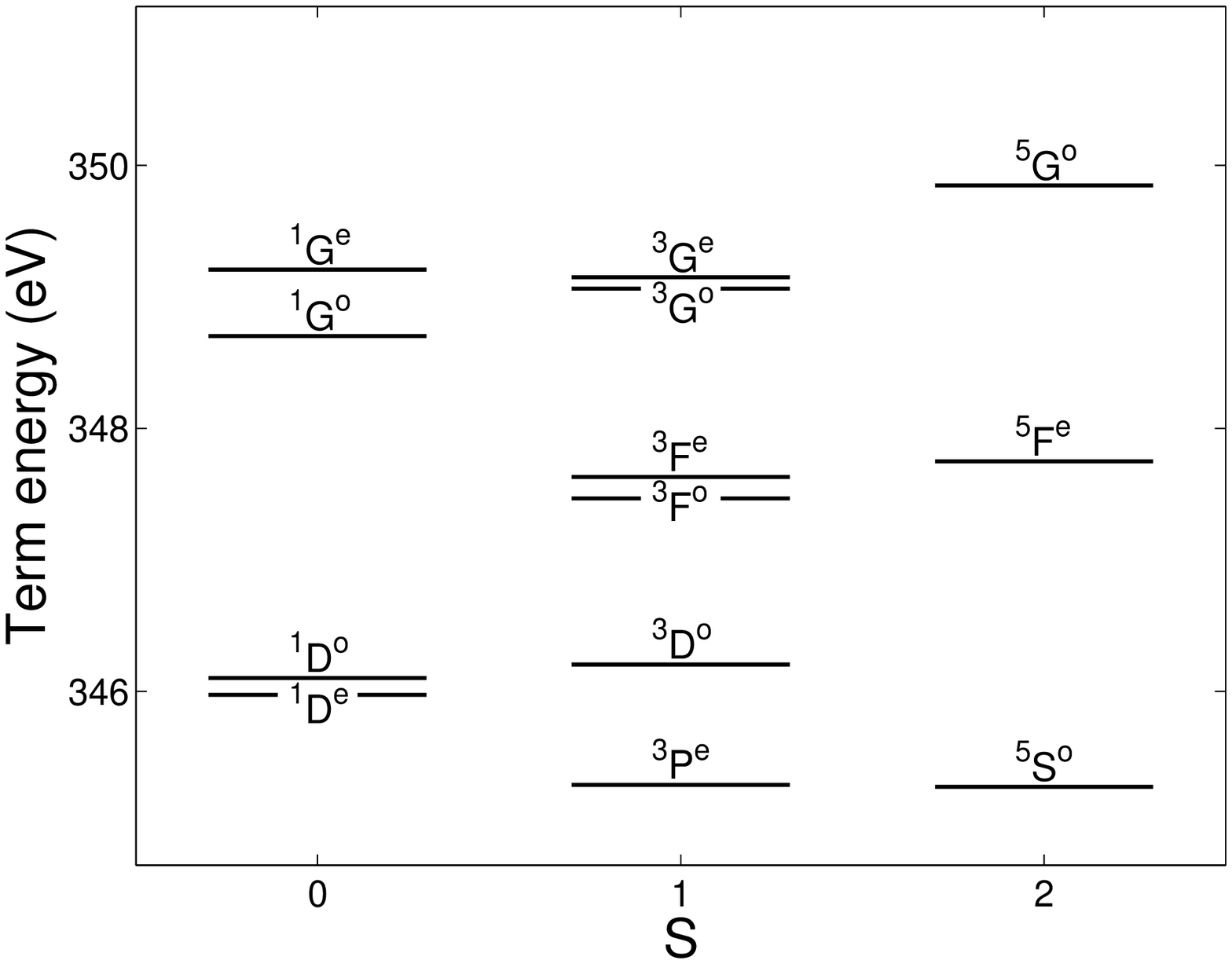}}
    \\
\subfigure[]{\includegraphics[scale=.4]{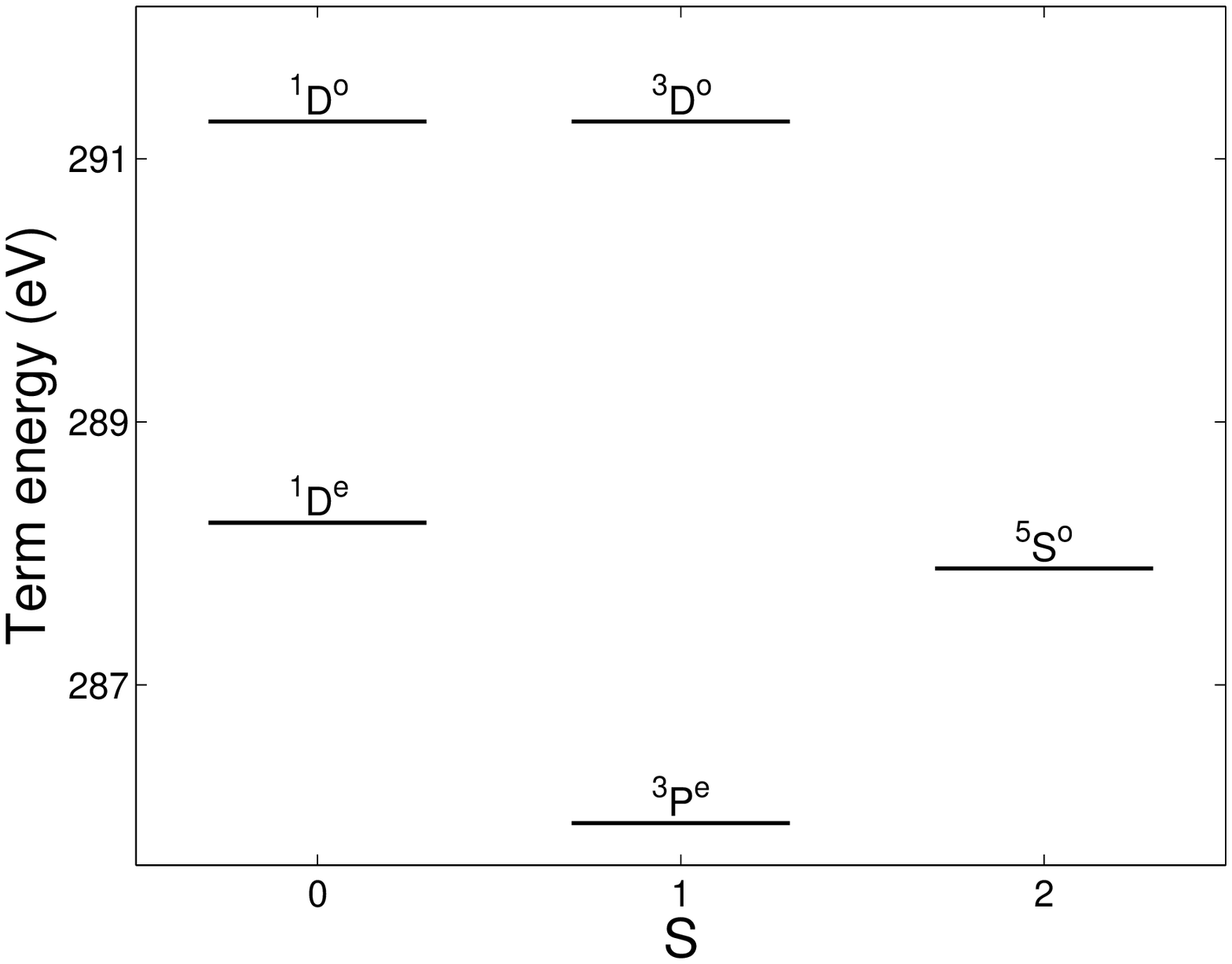}}
\end{tabular}\end{center}
    \caption{
Energy levels for pyramidal intra-shell states in Be. (a) $n=3$
states with $L\leq 4$. (b) $n=2$ states.}\label{Energy4dia}
\end{figure}

Our model provides an analytical expression for the wave function
describing quadruply excited intra-shell states of the
four-electron atom and gives a simple physical image of the atomic
states. To test the quality of the model, configuration-mixing
fractions are compared with \emph{ab initio} results. We expand
the ansatz in the spherical basis $
|\Psi_{R}^{LMM_I\pi,SM_s}\rangle
=\sum_{\bar{l}\bar{\mu}\bar{m}_s}c(\bar{l}\bar{\mu}\bar{m}_s,LMM_I\pi,SM_s)
\prod_{i=1}^4 |l_i\mu_im_{si} \rangle,$ ($\bar{p} =
(p_1,p_2,p_3,p_4), p=(l,\mu,m_s)$) with
$|\Psi_{R}^{LMM_I\pi,SM_s}\rangle$ being normalized such that
$\sum_{\bar{l}\bar{\mu}\bar{m}_s}|c(\bar{l}\bar{\mu}\bar{m}_s,LMM_I\pi,SM_s)|^2=1$.
The configuration-mixing fractions specifying the $l$ mixing
between several multiplet states are defined as
\begin{equation}\label{mixfrac4}
P_{\bar{l}_c}^{LM_I\pi
S}\equiv\sum_{\bar{l}\in\bar{l}_c}\sum_{\bar{\mu}\bar{m}_s}|c(\bar{l}\bar{\mu}\bar{m}_s,LMM_I\pi,SM_s)|^2\;,
\end{equation}
where $\bar{l}_c=(l_1l_2l_3l_4)_c$ denotes the $l$-configuration
regardless of permutation, e.g.,
$(sppp)_c=\{sppp,pspp,ppsp,ppps\}$. For $n=2$ the pyramidal states
are shown in Fig.\ref{Energy4dia}(b). The mixing fractions for
$^3$P$^\text{e}$ and $^1$D$^\text{e}$ are 0.9 $\text{s}^2
\text{p}^2$ and 0.1 $\text{p}^4$. The other terms have only a
single configuration ($\text{s}\text{p}^3$). Table \ref{5S} shows
a comparison between the mixing fractions predicted within the
model and from multiconfigurational Hartree-Fock calculations
\cite{KN}. The model not only predicts the trend in the mixing
fractions, but actually compares very well with the
configurational calculations. The convincing agreement in the
predictions shows that our four-electron model accounts for the
strong electron-electron correlations within the quadruply exited
intra-shell states. Table \ref{mixing4} shows configuration-mixing
fractions for all terms with $L\leq 4$ within the $n=3$ shell for
which the pyramidal shape is accessible~\cite{Bao96,Bao98}.
\begin{table}[!h]
  \centering
  \caption{Configuration-mixing fractions for pyramidal states in Be for the $n=3$
  shell with $L\le4$.}
 \begin{tabular}{c c r r r}
   \hline
   \hline
   Term & Configuration & $S=0$ & $\phantom{1}S=1$ & $\phantom{1}S=2$\\
   \hline
   ${}^{2S+1}\text{S}^\text{o}$ & sp$^3$ & & &0.9375\\
   & spd$^2$ & & &0.0625\\
   \vspace{-3.1mm}\\
   ${}^{2S+1}\text{D}^\text{o}$ & s$^2$pd & 0.5265 & \phantom{1}0.1529\\
   & sp$^3$ & 0.2632 & 0.6878\\
   & p$^3$d & 0.1504 & 0.0983\\
   & spd$^2$ & 0.0501 & 0.0546\\
   & pd$^3$ & 0.0097 & 0.0065\\
   \vspace{-3.1mm}\\
   ${}^{2S+1}\text{F}^\text{o}$ & s$^2$pd & & 0.6490\\
   & spd$^2$ & & 0.2163\\
   & spd$^2$ & & 0.1217\\
   & pd$^3$ & & 0.0129\\
   \vspace{-3.1mm}\\
   ${}^{2S+1}\text{G}^\text{o}$ & p$^3$d & 0.8010 & 0.5659\\
   & spd$^2$ & 0.1424 & 0.4024 & 1.0000\\
   & pd$^3$ & 0.0566 & 0.0316 & \\
   \vspace{-3.1mm}\\
   ${}^{2S+1}\text{P}^\text{e}$ & s$^2$p$^2$ & & 0.6082\\
   & sp$^2$d & & 0.2027\\
   & p$^4$ & & 0.1521\\
   & p$^2$d$^2$ & & 0.0174\\
   & s$^2$d$^2$ & & 0.0135\\
   & sd$^3$ & & 0.0058\\
   & d$^4$ & & 0.0003\\
   \vspace{-3.1mm}\\
   ${}^{2S+1}\text{D}^\text{e}$ & s$^2$p$^2$ & 0.6204\\
   & p$^4$ & 0.1551\\
   & sp$^2$d & 0.1477\\
   & p$^2$d$^2$ & 0.0443\\
   & s$^2$d$^2$ & 0.0295\\
   & sd$^3$ & 0.0022\\
   & d$^4$ & 0.0008\\
   \vspace{-3.1mm}\\
   ${}^{2S+1}\text{F}^\text{e}$ & sp$^2$d & & 0.7829 & 0.9643\\
   & s$^2$d$^2$ & & 0.1392\\
   & p$^2$d$^2$ & & 0.0706\\
   & sd$^3$ & & 0.0058 & 0.0357\\
   & d$^4$ & & 0.0016\\
   \vspace{-3.1mm}\\
   ${}^{2S+1}\text{G}^\text{e}$ & sp$^2$d & 0.4165& 0.7692\\
   & s$^2$d$^2$ & 0.2777\\
   & p$^2$d$^2$ & 0.2752 & 0.1795\\
   & sd$^3$ & 0.0278 & 0.0513\\
   & d$^4$ & 0.0028\\
   \hline
\end{tabular}
\label{mixing4}
\end{table}
The mixing fractions are seen to be strongly spin dependent. As a
consequence of Pauli statistics some configurations are even
forbidden for accessible terms. Furthermore, there is a notable
difference between the mixing fractions for quadruply excited
states and those for triply excited states~\cite{MMPRA}: Contrary
to triply excited states, the mixing fractions for quadruply
excited intra-shell states do not depend on $M_I$. This reflects
the three dimensional symmetry of the regular tetrahedron
(spherical top) compared to two dimensional symmetry of an
equilateral triangle (symmetric top).

The energy of a state within the symmetric rotor model is given by
$\langle \Psi_{R}^{LM_I\pi S}\left| \hat{H}
\right|\Psi_{R}^{LM_I\pi S}\rangle$, i.e.,
\begin{equation}
E_{n}^{LM_I\pi S}= 4E_n+\sum_{i<j}\left\langle\Psi_{R}^{LM_I\pi
S}\left|\frac{e^2}{r_{ij}}\right|\Psi_{R}^{LM_I\pi
S}\right\rangle\;,\label{Erotorchap3}
\end{equation}
with $r_{ij}=|\vec{r_i}-\vec{r_j}|$. Here $E_n$ is the Bohr energy
and the last term is the electron-electron interaction energy. The
equality follows from the fact that the Stark states are
eigenstates of the single-particle part of the Hamiltonian of
Eq.~\eqref{Hamil}.

Figure \ref{Energy4dia} shows the energies predicted within our
model for the pyramidal states within $n=2$ and for $L\leq4$
within the $n=3$ shell. The energy increases with increasing $L$.
Furthermore, we see the tendency that the energy increases with
increasing spin, $S$, contrary to the behavior in ground states
configurations where the term with the largest possible value of
$S$ for a given $L$ has lowest energy (Hund's rules). Finally, the
energy levels split up for different parity since different parity
requires different configurations. We note that if the system is
considered from a molecular viewpoint the $M_I$-independence of
the configuration-mixing fractions is expected since the
rotational energy of the spherical top molecule reads
$E=\frac{1}{2I}L(L+1)$, with $I=\sum_imr_i^2$ being the moment of
inertia of the molecule. However, the $L$ scaling does not follow
this relation and it is clear from Eq.~\eqref{Erotorchap3} that
rotationallike structure in excited atomic states is not
associated with the kinetic energy term as in a molecule, but is
exclusively due to the electron-electron interaction. The term,
$\left\langle\Psi_{R}^{LM_I\pi
S}\left|\frac{e^2}{r_{ij}}\right|\Psi_{R}^{LM_I\pi
S}\right\rangle$, scales linearly with the nuclear charge, $Z$,
due to the scaling of the matrix element of $\frac{1}{r_{ij}}$.
Contrary, the spherical top model predicts a $Z^2$ scaling of the
energies due to the scaling of the moment of inertia.

In summary, we constructed an analytical wave function for
quadruply excited states corresponding to the case where the four
excited electrons are in the corners of a pyramid. By comparison
with {\it ab initio} calculations, we confirmed the accuracy of
the model and, hence, the appealing physical image of the
correlated motion of the four electrons. We predicted
configuration mixing fractions for a number of states and
generally found that a labeling of states in terms of single
configurations is insufficient. Finally, we calculated the
spectrum of pyramidal states within the second and third principal
shell and compared with the expectations obtained when viewing the
system as a molecularlike spherical rotor.

We propose to address pyramidal states by dipole allowed light
absorption transferring  a fraction of the Be ground state
population to the singly excited Be(1s1s2s2p$^1$P$^\text{o}$)
state by resonant laser excitation (5.277 eV). A synchrotron or a
free electron laser source could then subsequently scan through
the region of quadruply excited pyramidal states of
$^1$D$^\text{e}$ symmetry. A similar two-color technique was
successfully applied on triply excited states in Li
\cite{Cubaynes96}.

\begin{acknowledgments}
  LBM is supported by the Danish Natural Science Research Council (Grant
No. 21-03-0163).
\end{acknowledgments}


\end{document}